# Vibration by relativistic effects.

Enrique Ordaz Romay[1]

*Facultad de Ciencias Físicas, Universidad Complutense de Madrid*

## Abstrac.

Relativity, time reversal invariance in mechanics and principle of causality can be in the bases of a type of vibration of the extensive objects. It is because, the detailed analysis of the relativistic movement of an extensive body entail that all the objects must have inherent a vibratory movement to their own size. Such effect does not happen when it works with point particles thus is not stranger who happens unnoticed in the traditional studies.

Also we can find relation between the form of vibration of the extensive objects and the energy that calculates by quantum considerations.

---
[1] eorgazro@cofis.es

# Introduction.

Usually, the mathematical developments and models that are used in classic and relativistic physics are supported in the concept of material point. This technique is very useful in many cases, but in others it can suppose a problem because it prevents to determine some exclusive phenomena of the objects that have extension.

In the detailed analysis of the relativistic movement of a body we observed that relativity, with the time reversal invariance in mechanics and the principle of causality entail to that all the extensive objects must have inherent a vibratory movement which manifestation is not easy to understand from the point of view of the physics of material points.

# The Lorentz contraction.

In the classic relativistic study of the movement of extensive bodies in inertial frames of reference, is deduced the phenomenon known like Lorentz contraction. This phenomenon is studied of the following way:

Let $s'$ a inertial reference frame in which the extensive object is in rest. Be also, the movement of the extensive object respect to the referential inertial frame $s$, according to the direction of the $X$ axis, with a speed $v$, being "*begin*" and "*end*" the ends of the extensive object throughout that direction. In this case, the rest length of the object is $l_0 = x'_{end} - x'_{begin}$, while in the frame in which the object is in movement the length is $l = x_{end} - x_{begin}$.

According to Lorentz transformations in the change of a one inertial frame to another, it obtained (being $c$ the light speed):

$$x'_{begin} = \frac{x_{begin} - vt_{begin}}{\sqrt{1 - \frac{v^2}{c^2}}} \qquad x'_{end} = \frac{x_{end} - vt_{end}}{\sqrt{1 - \frac{v^2}{c^2}}}$$

In order to measure the object length *l*, the light rays that beginning in "*begin*" and "*end*" of the object must arrive simultaneously to the observation point in the frame *s*. For it, $t_{begin} = t_{end}$. Replacing it has left:

$$l_0 = x'_{end} - x'_{begin} = \frac{x_{end} - x_{begin}}{\sqrt{1 - \frac{v^2}{c^2}}} = \frac{l}{\sqrt{1 - \frac{v^2}{c^2}}}$$

Consequently, when one object with rest length $l_0$ is moving in an inertial frame with a *v* constant velocity, the length *l* measurein this frame will be:

$$l = l_0 \cdot \sqrt{1 - \frac{v^2}{c^2}}$$

This expression is known like Lorentz contraction.

This development is totally correct from an aprioristic point of view. That is to say, if we suppose that, independently of the causes that have originated the movement of the body respect to the inertial frame *s*, the object is moving with constant speed in the frame.

Nevertheless, the fact that an extensive object can be moved as a whole with constant speed in an inertial frame is a questionable hypothesis because, in relativity, the interactions do not travel instantaneously between two points, but that take a time in arriving from a point at another one because the maximum velocity is the speed of the light. Consequently, all the bodies are deformable and all movement necessarily deforms to the bodies.

This fact forces to ask: Can exist in relativity no-punctual objects that move like a whole in an inertial frame?

Let us suppose that we have an extensive body in rest and at a moment we applied an interaction that puts it in movement. Passed a smaller time than the lapse taken by the light in crossing all the length of the body, this will be become deformed because some parts of him are in movement and others are in rest. Only when the light has crossed the totality of the body this fulfils the conditions of the Lorentz contraction.

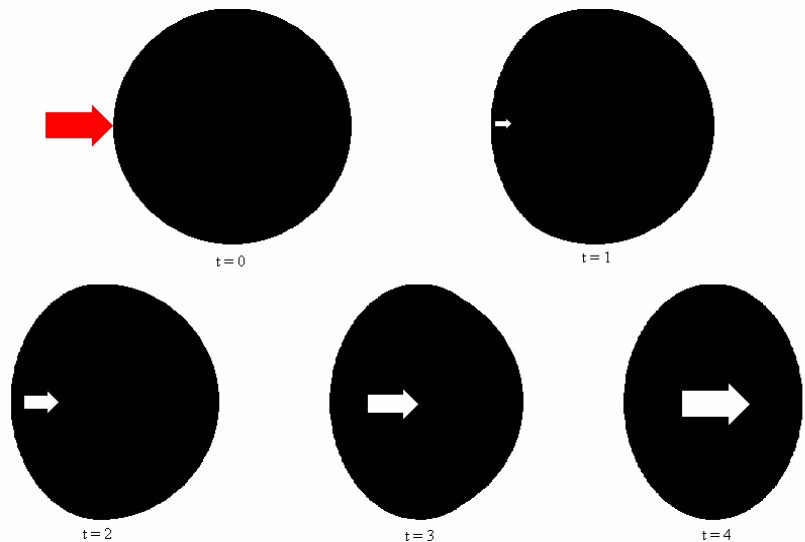

Nevertheless, once this deformation finishes, that is to say, when the light has already surpassed the totality of the extension of the object, Is its final state deformed (Lorentz contraction) or maintains an internal expansion-contraction movement in him like an internal vibration existed?

In order to make an approach to this problem, in the present work we are going to study as it is the form of the movement of a simple system formed by an precise object that hits against two united precise particles through a holonomic constraints. Through this scheme and its properties we will be able to generalize the form in which the movements of the extensive bodies and their properties take place.

## The collision of a particle against a punctual particle pair.

We are going to analyze the classic relativistic problem of the collision of a no-polar punctual particle to against punctual particle pair united by a holonomic constraints $L$ that in rest has a length $l$ (that we will call in brief "pair"). In this case the sequence of the collision comes illustrated in the graphic 1.

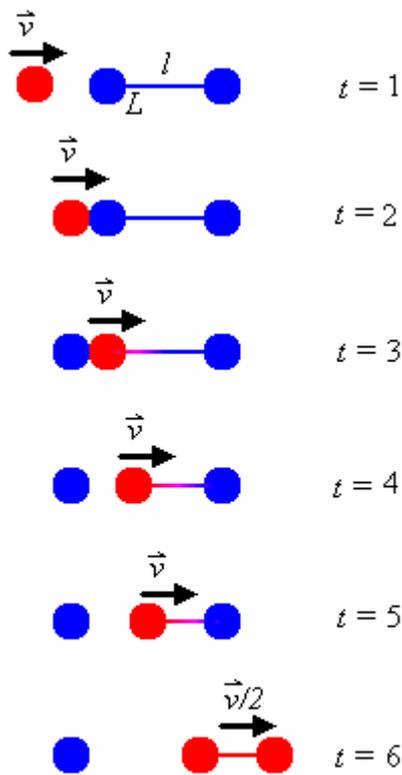

Graphic 1

In this graph, for $t = 1$ we see, in red the incident punctual particle (we showed it like a small single sphere so that its position is appraised well) that moves with speed $v$ towards the pair formed by two identical in mass and united punctual particles by means of a constraints $L$.

For $t = 2$ takes place the perfectly elastic collision. For $t = 3$ the movement to the first particle of the pair is transmitted, being the incident particle in rest.

For $t = 4$ and 5 it is appraised that, the transmission of the movement is not instantaneous, because it travels at an equal terminal velocity at the speed of the light, since the first particle of the pair is put in movement until it is put in movement second spends a time during which, constraints $L$ diminishes its length.

For $t = 6$, finally, the third punctual particle is put in movement so that, the speed of the pair, according to the law of conservation of the linear moment, is of $v/2$ and the final length of the segment, according to the known phenomenon as Lorentz contraction is $l = l_0 \big/ \sqrt{1 - v^2/4c^2}$ .

## The apparent violation of the time reversal or of the causality.

Nevertheless, the previous scheme has a serious disadvantage: Apparently the conservation of the time reversal or the causality principle is violated.

Let us determine the first two following definitions:

- The time reversal invariance says that, in mechanics, if a phenomenon occurs in the nature, when changing the direction of the time such phenomenon is also physically possible.

- On the other hand, the causality principle says that all events are been from those events that precede temporarily to it with a positive interval.

Having this in account we analyze the problem of the time reversal of the previous collision. In a temporary investment the previous graphs must follow one another in inverted order. That is to say, consecutively in graphic 1 from $t = 6$ to 1. Now we will represent in Graphic 2 from $t = 1$ to 6.

In these conditions we were now with a causal problem: beginning in graphic 2, in $t = 2$ we would be that, without apparent reason, the second particle of the pair stops its movement, transmitting it to the first particle. In $t = 3$ and 4 the shutdown of the second particle is transmitted through the constraints until the particle first.

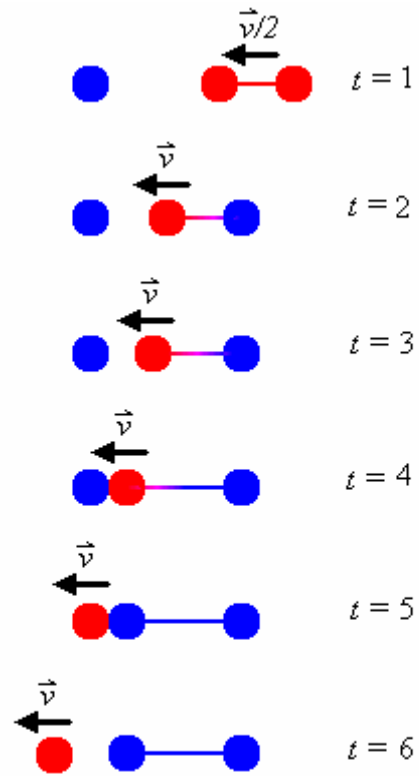

Graphic 2

In $t = 5$ the movement is transmitted from the pair until the particle in rest and, in $t = 6$, the particle moves away of the pair.

It is possible to be appreciated that in this process $t = 2$ supposes one pretends violation of the causality principle, because, the second particle of the pair is stopped without a previous cause.

Due to this opportune and no-causal shutdown of only one of particles of the pair, produce an increase of the length of the constraints. This event is undoing the Lorentz contraction while the shutdown of one individual particle of the pair is transmitted, through the constraints. We say that the shutdown is opportune because it makes agree the moment of length in rest of the tie with the collision of this particle of the pair against the free particle.

Now we entered the true nature of the created problem:

- If special relativity preserves the causality, the inverse sequence is not possible (because in graphic 2, $t = 2$ would violate the causality principle). That is to say, special relativity would violate the time reversal.

- If special relativity preserves the time reversal the inverse sequence must be possible and special relativity would violate the causality principle.

How must it be an elastic collision according to the laws of the causality for initial conditions like those of the previous process?. At first the speed of particles of the pair would have to stay constant until the incident particle of the pair hit against the free particle. In this case, the graphical representation must have the form of the graphic 3.

This representation has serious disadvantages, all derivatives from the same cause: As the incident particle of the pair hits first against the free particle (graphic 3, $t = 2$) yielding its momentum ($t = 3$) and therefore stopping itself, the second particle of the pair cannot yield its moment to the free particle without violating the causality, that is to say, the passage from the $t = 3$ to 5 is impossible.

Another one of the problems we found it carrying out the time reversal of this last situation. Passing from the $t = 5$ to the 3 is causally impossible.

Consequently, the evident form of elastic collision between a pair and a punctual particle, if it happens simply by direct bonding of one of particles

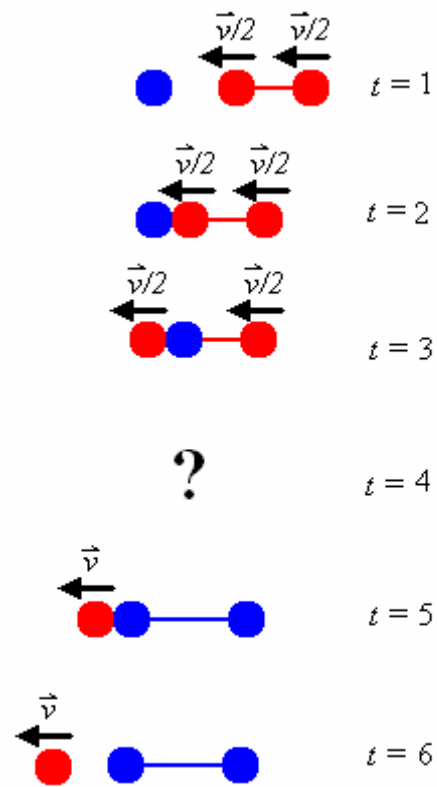

Graphic 3

of the pair, violates the causality, again.

## The vibratory movement of the extensive objects.

We are going to return to raise the aligned direct collision of a particle with the pair. Now we will analyze separately the speeds of each one of punctual particles of the pair and their mutual interaction and will continue our observations during a longer time period. The corresponding graph is the graphic 4.

When we analyzed the movement of each one of the parts of the pair, we are forced to consider the constraint as another part of the deformed object and its deformation is due to maintain the causal consistency and to allow the conservation of the laws of the mechanics before the time reversal.

The resulting analysis of this elastic collision is the following one: the pair, like a whole, besides to move its center of masses at a certain speed, maintains a vibration movement

This vibration is a necessary result phenomenon of the elastic collision with an extensive object when we consider the relativistic phenomena such as the not instantaneous transmission of the interactions.

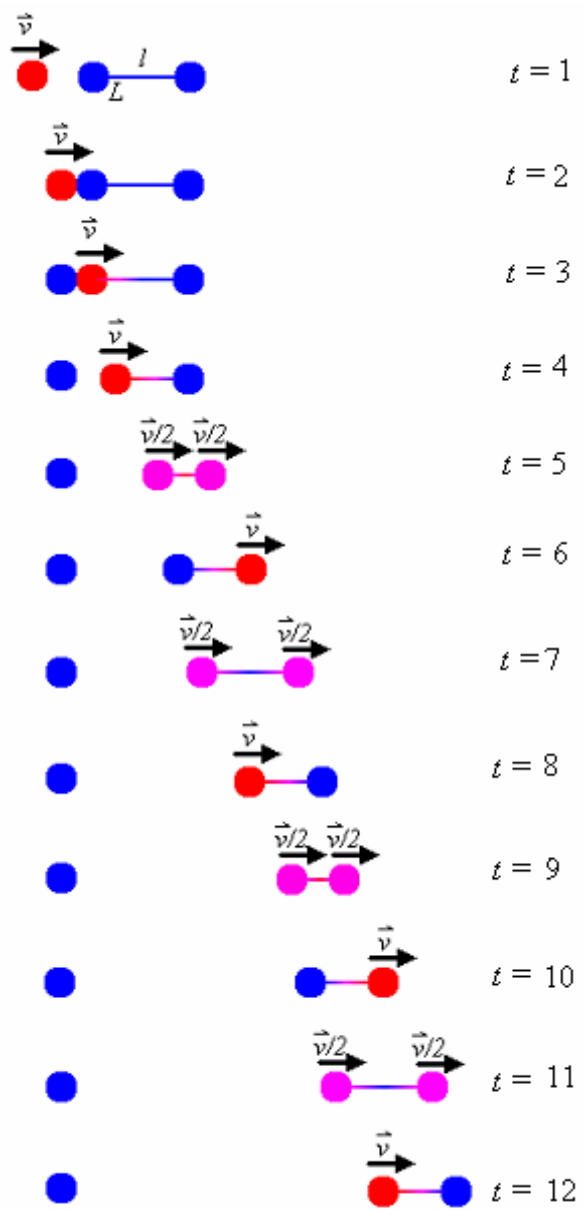

Graphic 4

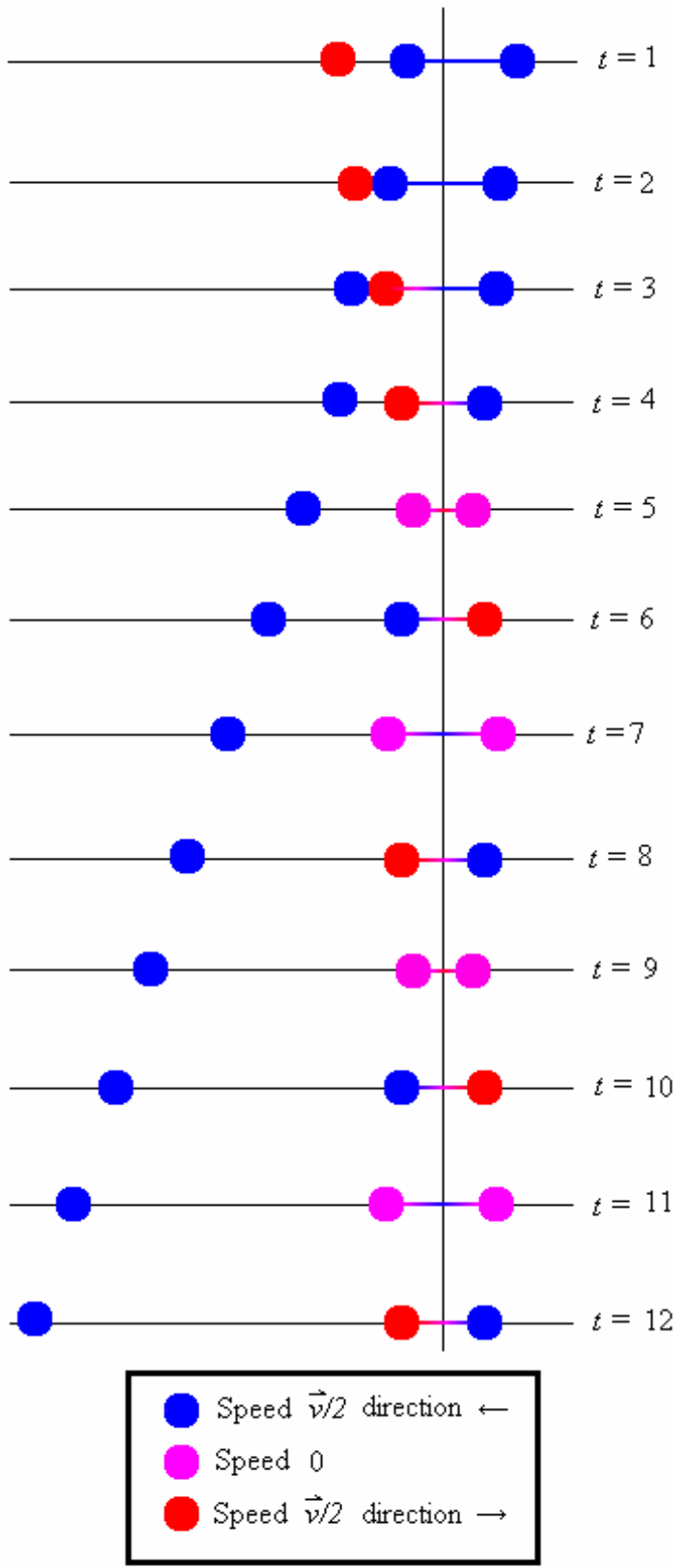

Graphic 5

Up to here, we have used the frame in which the incident punctual particle is in rest after the collision. The vibration of the pair after the collision is appraised better in a frame that moves in the direction of the initial incident particle with an equal speed to the center of mass.

The graphic 5 corresponding to this reference frame (we remembered to apply the Lorentz contraction due to the movement of the frame).

In these graphic it is seen that the collision between the pair and the punctual particle task, in addition to the interchange of lineal momentum, a vibration in the pair.

The process of "flux" of kinetic energy through the constrain between two punctual particles of the pair is appraised now better. This effect is appraised better thanks to the colour that we have applied on the constrain. Thus we found that the laws of the causality as of the time reversal are conserved as much.

The form of the movement of the pair after the collision with a punctual particle is, by all the sight up to here, a vibration.

In general, all movements are produce by the linear momentum transmission, or by the collision of a particle or the interaction with a field (the crash "emission-absorption" with particles of such field). Because, in any case, the transmission of the movement within the pair does not take place of instantaneous way, after a collision or interaction, a pair of punctual particles must be in vibration state.

Consequently, all pair of punctual particles in movement will have to show a movement of vibration due to the transmission of such movement by the constraint that unites to them.

Supposing the simplest case, in our analysis there are assumption, in all the cases, that before the collision the particles of the pair were not in movement. Nevertheless, a priori this idea is not correct by several reasons. At first, the own existence of the pair implies the interaction between both particles of the pair and since the interaction cannot travel of instantaneous form through the length of the constrain, the formation of the pair forces a vibratory movement.

But on the other hand, if a pair of particles united by a constrain maintained a state of no vibration, this would lead until a problem with the principle of relativity because the laws of the physics would be different in two inertial frames of reference.

Let us suppose that in an inertial frame of reference a pair (in movement or rest) did not vibrate. In order to conserve the causality and the time reversal in another inertial frame in movement with the first, the pair would have to vibrate. Nevertheless, as it is deduced of the Lorentz transformations, the frequency of vibration ($v$) of a body in the change of a inertial frame to another one with speed $v_s$ verifies the expression ($\omega$ angular frequency):

$$v' = \frac{v}{\sqrt{1 - \frac{v_s^2}{c^2}}} \Rightarrow v' = v \cdot \gamma_s \quad (1)$$

$$\omega = 2\pi \cdot v \Rightarrow \omega' = \omega \cdot \gamma_s$$

That is to say, the vibration frequency diminishes with the speed, tending to zero when the speed tends to *c*.

Consequently, it is impossible that a pair vibrates in a frame and does not vibrate in another one.

The only one solution to conciliate relativity, time reversal and causality, as well as the mechanic formation of the pairs (by mutual interactions between particles) consists of considering that all pair of particles and therefore all extensive body have a natural state of vibration.

We will call factor of vibration RTC to the no-contradiction between relativity, the time reversal and the causality like cause of the state of fundamental vibration of an extensive object.

## Calculus of the vibration parameters of a pair.

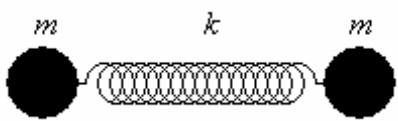

Let us suppose an approach for the vibration problem by factors RTC to the movement of a pair of particles with rest mass $m_0$ united by a wharf with elastic constant *k*.

In this system the vibration angular frequency of the pair in the first relativistic correction (for particle speed $v_p \ll c$) is $\omega = \dfrac{1}{\gamma_p}\sqrt{\dfrac{k}{m_0}}$ being $\dfrac{1}{\gamma_p} = \sqrt{1 - \dfrac{v_p^2}{c^2}}$.

Applying (1) to this expression we have left the relativistic form of the vibration frequency of the pair: $\omega = \dfrac{\gamma_s}{\gamma_p}\sqrt{\dfrac{k}{m_0}}$

We calculated the energy of this system applying the quantum energy equation for an object that vibrates: $E = (n + \frac{1}{2}) \cdot \hbar\omega$ being $\omega$ the vibration angular frequency in the rest frame and $n$ a natural number. Replacing we have left:

$$E = \hbar(n+\tfrac{1}{2})\frac{\gamma_s}{\gamma_p}\sqrt{\frac{k}{m_0}}$$

On the other hand, the energy of the system can be calculated by the Einstein expression of the energy: $E = \gamma_t \cdot M_0 \cdot c^2$, being $M_0$ the mass in rest of complete system, which in our case is $2 \cdot m_0$ and $\gamma_t = \dfrac{1}{\sqrt{1-\dfrac{v_t^2}{c^2}}}$ with $v_t = \dfrac{v_p + v_s}{1+\dfrac{v_p v_s}{c^2}}$ the total speed of the particle.

Replacing and clearing $k$ and $\omega$ we have left:

$$k = 4\frac{c^4}{\hbar^2(n+\tfrac{1}{2})^2}\frac{\gamma_t^2 \gamma_p^2}{\gamma_s^2}m_0^3$$

$$\omega = 2\frac{c^2}{\hbar(n+\tfrac{1}{2})}\gamma_t m_0$$

that corresponds to the value of the elastic constants of the constrain of the pair of mass $2 \cdot m_0$.

From this last expression it is deduced that the vibration frequency is quantified. For $n = 0$ we have the traditional value of the frequency of $\omega = 2\dfrac{c^2}{\hbar}\gamma_t m_0$

The elastic energy of the system is: $E_{elastic} = \dfrac{1}{2}k \cdot (Amplitude)^2$. Being the amplitude of the vibratory movement the difference between $l$ and $l_0$, then: $E_{elastic} = \dfrac{1}{2}k \cdot l_0^2(\gamma_p - 1)^2$.

In this last expression we replaced the value found for $k$:

$$E_{elastic} = 2\frac{c^4}{\hbar^2(n+\tfrac{1}{2})^2}m_0^3 \cdot l_0^2 \frac{\gamma_t^2 \gamma_p^2(\gamma_p - 1)}{\gamma_s^2}$$

The elastic energy is invested in movement, that is to say, in kinetic energy whose relativistic expression is $E_{kinetic} = (\gamma - 1) \cdot M_0 \cdot c^2$. Replacing in the expression of the elastic energy and leaving the constants in the right member we have left:

$$m_0 l_0 \frac{\gamma_s}{\gamma_t \gamma_p (\gamma_p - 1)} = \sqrt{2} \cdot \frac{\hbar}{c}(n + \tfrac{1}{2})$$

When we are in the frame in which the set of the pair, although vibrating, its centre mass is in rest (that is to say, $\gamma_s = 1$ y $\gamma_t = \gamma_p$ ), for $v_p \ll c$ we have left ($\frac{1}{\gamma_p^2 (\gamma_p - 1)^{1/2}} \approx \sqrt{2} \frac{c}{v_p}$):

$$m_0 l_0 \frac{1}{\gamma_p^2 (\gamma_p - 1)^{1/2}} = \sqrt{2} \cdot \frac{\hbar}{c}(n + \tfrac{1}{2}) \Rightarrow m_0 l_0 \approx \frac{\hbar}{c^2}(n + \tfrac{1}{2}) \cdot v_p \qquad (2)$$

These last expressions indicate to us that the product $m_0 l_0$ of a vibrating pair is quantized and its value depends on the speed of vibration.

From this last expression we obtain the minimum value for the product $m_0 l_0$ of a pair. Making $n = 0$, for the product $m_0 l_0$ we have left:

$$m_0 l_0 \approx \frac{\hbar}{2c^2} \cdot v_p$$

## Can RTC vibration offer an explanation to the quantum phenomenon?

As well it is known, the quantum mechanics part from the supposition of which the physical objects and very especially the subatomic particles behave simultaneously like wave and corpuscle. The mathematical model of this behavior is implemented in the wave function.

It is possible to be observed that all the known subatomic particles, with mass, although their dimensions are very small they are extensive objects. Even the electron, the smaller particle with mass, has a classic radius of $2.818 \cdot 10^{-15}$ meters. Its value is not excellent because this numerical result is not obtained by totally quantum-relativistic analysis. Nevertheless, it is excellent the fact that it has an extension.

Thus, all the extensive subatomic particles are subject to vibrations by effects RTC. That is to say, RTC effects imply a vibration in all the macroscopic objects as much as microscopic, like the quantum phenomenon.

In our previous development, by facility, we have initiated from the quantum expression of the energy for a state of vibration with quadratic form for potential energy. Nevertheless, the expression (2) can be deduced by classic mechanical analysis relativistic in the next form.

In a vibratory movement the speed of the particle is proportional to the product of the frequency by the displacement: $v_p \propto \omega l$.

The speed in the vibratory movement is limited and out of phase in 90º respect to the displacement therefore, for n periods we obtain $(n + ½) \cdot v_p \approx \omega l$.

Multiplying both members by the constant $1/c^2$ and multiplying and dividing the second member by the mass we obtain

$$\frac{1}{c^2}(n + ½) \cdot v_p \approx \frac{\omega}{mc^2} ml .\tag{3}$$

From relativity, the term $mc^2$ represents the energy of the system. On the other hand, since the energy that can emit a vibrant pair is finite, it is to say: $\int_0^\infty E(\nu)d\nu < +\infty$, then, a value exists of $\nu$ that it makes maximal to the energy. Whichever greater it is the frequency of vibration (and therefore of radiation) of the pair, greater is the energy of the system, being linear the relation between both, for small speeds of the pair particles: $E_{max} \propto \nu_0$.

This reasoning is altogether experimental agreement with the law of the displacement of Wien immediately than to observe than $E = {}^3/_2\ kT$ and that $\lambda \nu = c$ (being here $k$ the constant of Boltzmann and $T$ the temperature corresponding to the energy of the pair). Replacing, we found the expression of the law of the displacement: $\lambda_0 T = constant$.

Then, the energy of the vibrant pair will be proportional to the frequency of vibration of the pair, is to say $mc^2 \propto \nu$. We called $h$ to the proportionality constant and replacing in the expression (3) we were able to obtain the expression (2), with no need to use quantum postulates.

A concrete case of vibrant pair that it fulfills these considerations is the mesons particles. These are composed by two quarks united by constrain formed by the strong nuclear force. This is indeed the scheme studied in the present work. Consequently the mesons must have a form of characteristic vibration similar to the obtained one in the previous expressions.

In the present work we have been centred in the study of the vibrant pairs. Nevertheless, the vibrations by factors RTC must happen in any extensive bodies. For this reason, these vibrations can justify the quantum phenomena.

## Conclusions

1. All extensive body, due to factors RTC, is in vibration state.
2. Applying the quantum considerations it is observed that the vibration energy and the product $m_0 l_0$ are quantized.
3. Theoretical justification exists to associate the phenomena of vibration by effects RTC with the quantum phenomenon.